# Anomalous quantum criticality in the electron-doped cuprates

P. R. Mandal, Tarapada Sarkar, and Richard L. Greene

*Center for Nanophysics & Advanced Materials and Department of Physics, University of Maryland, College Park, Maryland 20742, USA.*

**In the physics of condensed matter, quantum critical phenomena and unconventional superconductivity are two major themes. In electron-doped cuprates, the low upper critical field allows one to study the putative QCP at low temperature and to understand its connection to the long standing problem of the origin of the high-Tc superconductivity. Here we present measurements of the low temperature normal state thermopower (S) of the electron-doped cuprate superconductor $La_{2-x}Ce_xCuO_4$ (LCCO) from *x*=0.11 to 0.19. We observe quantum critical S/T versus $\ln(\frac{1}{T})$ behavior over an unexpectedly wide doping range *x* = 0.15 - 0.17 above the putative QCP (x=0.14) with a slope that scales monotonically with the superconducting transition temperature. The presence of quantum criticality over a wide doping range provides a new window on the criticality. The thermopower behavior also suggests that the critical fluctuations are linked with Tc. Above the superconductivity dome, at *x*=0.19, a conventional Fermi-liquid S $\propto$ T behavior is found for T $\leq$ 40 K.**

**Introduction**

A quantum critical point (QCP) arises when a continuous transition between competing phases occurs at zero temperature. The existence of a QCP has been suggested in a variety of exotic materials, in particular under the superconducting dome in high-Tc copper oxides (cuprates) [1]. In strongly correlated materials displaying antiferromagnetic (AFM) order, such as heavy fermions, cuprates, and iron pnictides, quantum criticality is an important theme for understanding the low temperature physics and the superconductivity. In these materials it is believed that quantum fluctuations influence the physical properties over a wide temperature region above quantum critical point. In this region the system shows a marked deviation from conventional Landau Fermi-liquid (FL) behavior. The superconductivity (SC) in the cuprates may be governed by proximity to a QCP, although exactly how is still a mystery in spite of many years of intense research on these materials [1-4]. In hole-doped cuprates, a QCP has been found to be associated with the disappearance of the pseudogap phase [5-7], a phase of unknown origin. The electron-doped cuprates have a less complex doping phase diagram and a much lower magnetic critical field [8], which allows the T→0 K normal state to be studied over the entire phase diagram. The absence of pseudogap physics, and other unidentified competing phases, allows the QCP to be attributed to the disappearance of AFM as doping is increased away from the Mott insulating state at $x = 0$ [9]. However, the relation between quantum criticality and the normal state behavior of the *n*-type (and *p*-type) cuprates is still an important open question.

In the past, the transport properties of the *n*-type cuprates near the AFM QCP have been studied primarily by electrical resistivity and Hall Effect measurements [10] and Shubnikov-de Hass oscillations [11, 12]. These experiments, along with ARPES [13, 14], have given strong evidence for a Fermi surface reconstruction (FSR) at this AFM QCP, at a doping just above the optimal doping for superconductivity. In this Letter, we provide a surprising new insight on the quantum criticality via the first thermoelectric measurements in the field driven normal state of the electron-doped cuprate LCCO, for doping above and below the purported QCP. The temperature dependence of the thermopower at low temperatures provides a distinctive signature of quantum critical behaviour [15, 16].

**Results and discussions**

The normal state thermopower ($S$) measurements have been carried out from 4-80 K on $La_{2-x}Ce_xCuO_4$ thin films with doping from $x = 0.11$ to 0.17 in magnetic field of H>$H_{C2}$. Detailed information on the preparation/characterization of these films and the thermopower measurement technique is given in the SI [17], but also in [8]. Fig. 1a presents the data for $S$ (T) at normal state below 80 K plotted as S/T versus T for $x = 0.11$ to 0.17. Similar data is found for several films at each doping. For $x$=0.11 and 0.13, S/T displays a strong temperature dependence and below a temperature $T_{Smax}$ becomes increasingly negative. This shows that electrons dominate the low-temperature normal state thermopower for these dopings. The peak in S/T decreases from $T_{Smax} \approx$ 27 K for $x$=0.11 to $T_{Smax}\approx$15 K for 0.13. In Fig. 2 we show the data of Fig. 1a plotted as S/T vs ln T for the doping $x$= 0.15, 0.16 and 0.17. For all the doping, the low temperature behavior of S/T goes as ln (1/T), with a deviation away from this dependence at higher temperature.

The dramatic change in the sign and magnitude of S/T from overdoped to underdoped region at 4 K is consistent with the Hall effect [18], where the 4K value of $R_H$ is observed to change from negative for $x < 0.14$ to positive above $x > 0.14$. As shown in supplementary Fig. S2 the normal state Hall resistivity maxima, $T_{R_{Hmax}}$ (the temperature below which Hall coefficient starts to fall) and $T_{Smax}$ lie on the same line, which is the estimated FSR line, $T_{FSR}$. The $T_{FSR}$ separates the large, hole-like, Fermi surface region from the reconstructed FS surface, In the T- $x$ phase diagram the FSR occurs in the region where commensurate (π,π) SDW modulations have been detected by neutron diffraction in other electron-doped cuprates [9]. For LCCO, AFM order has been detected by both in-plane angular magnetoresistance [19] and µSR experiments [20]. Moreover, Quantum Oscillation [12] and ARPES [21] measurements have seen evidence for the reconstructed FS for $x < 0.14$. Further such measurements are needed to confirm the existence of the large hole-like FS for $x > 0.14$, as suggested by our thermopower and Hall experiments. Thus the experimental evidence to date suggests that there is an AFM QCP at $x = 0.14$ for LCCO.

One expects that fluctuations associated with this QCP at T= 0 K will impact transport (and other properties) at finite temperatures above the QCP [2]. The most studied of these transport properties is the non-FL resistivity ($\rho \sim T^n$, with $n<2$) at low temperatures [22, 23]. In addition, in some heavy fermion materials a non-FL logarithmic temperature dependence of the low temperature thermopower has also been observed near a magnetic QCP [24]. This behavior of S(T) has been interpreted to result from low energy quasi-two-dimensional (2D) spin fluctuations associated with the AFM QCP [15]. In this theory, the thermopower is given by

$$S = \frac{1}{e}\left(\frac{g_0^2 \mathcal{N}'(0)}{\epsilon_F \omega_S \mathcal{N}(0)}\right) T \ln(\omega_S/\delta) \tag{1}$$

where, $\mathcal{N}(0)$ is the density of states at the Fermi energy $\epsilon_F$, $g_0^2$ is the coupling between the electrons and the spin fluctuations, $and$ $\omega_S$ is the energy of the spin fluctuations. Here $\delta$ measures the deviation from the critical point, which depends on the experimental parameters like doping, pressure, or magnetic field that can be tuned to the critical point. When T is greater than zero, temperature dependent thermopower is given by $S/T \propto \ln(1/T)$ in proximity to the QCP. Away from the critical point the thermopower shows a crossover to a Fermi-liquid behavior $S/T \propto$ $constant$ as T decreases. As the critical point is approached, the quantum critical behavior $S/T \propto \ln(\frac{1}{T})$ can be observed in a limited $T$ range.

Our thermopower data shown in Fig. 2 are in qualitative agreement with the Paul and Kotliar theory [15] at least down to 4 K. Surprisingly, we find $S/T \propto \ln\left(\frac{1}{T}\right)$ over a wide doping range, not just at the QCP, with no sign of a low temperature deviation towards S/T being constant at any doping. This suggests an "anomalous quantum criticality" in LCCO with a quantum critical region from $x \geq 0.14$ to the end of the SC dome at $x_c$ ~0.175. Above $x_c$ we find conventional FL behavior $S/T \propto constant$ at $x$=0.19 [Fig. 1 (b)]. To better understand the anomalous critical behavior in LCCO we have reanalyzed our prior thermopower data of the electron doped cuprate, $Pr_{2-x}Ce_xCuO_4$ (PCCO) measured in the normal state [25]. Figures 2 present the temperature dependence of S/T vs ln T for PCCO down to 3 K with doping $x$= 0.16, 0.17 and 0.19 at 9 T. The S/T shows a ln (1/T) dependence down to the lowest measured temperature for all the doping and deviates from its linear ln (1/T) behavior at higher temperature. The comparison between the normal state thermopower in PCCO and LCCO shows that the slope $A_{TEP}$ for both materials scales monotonically with the change of $T_C$ for different doping as shown in Fig 3. So this behavior appears to be universal in the electron-doped copper oxides. Our data is also supported by the low temperature normal state resistivity behavior for LCCO, where for $x$=0.15, 0.16, and 0.17 the resistivity varies linearly with temperature down to 20 mK [22]. So the breadth of the critical region in LCCO (and PCCO) suggests that the physics in electron-doped cuprates is associated not with a QCP, but with a novel, extended quantum phase.

In Fig. 3 we show the coefficient $A_{TEP}(x)$ of the S/T logarithmic T dependence, obtained from fits to the low temperature regions with $S/T \propto \ln(1/T)$, as function of doping for both LCCO and PCCO. *A significant discovery of this work is that* $A_{TEP}(x)$ decreases with $T_c$ as $x$ increases and goes to zero at the doping where superceonductivity ends. From Eq. 1, $A_{TEP}(x)$ depends mainly on the strength, g, of the coupling between the electrons and the spin fluctuations. Therefore, the strength of this coupling appears to be directly linked to the electron pairing (and hence the magnitude of Tc) in the n-type cuprates.

Figure 1 (b) presents our thermopower data for a non-superconducting film of LCCO ($x$ = 0.19, i.e., beyond the superconducting dome). In a conventional Fermi-liquid, we expect the low temperature thermopower to follow [26]:

$$S = \frac{\pi^2}{3}\frac{k_B}{e}\frac{T}{T_F} \qquad (2)$$

We use our data to estimate the Fermi temperature and the Fermi energy ($T_F = \epsilon_F/k_B$) from the slope of S vs T. We find $\epsilon_F \approx 10{,}000$ K$^{-1}$, which is in agreement with prior estimates for n-type cuprates from other experiments [27]. The abrupt change in low temperature thermopower behavior from non-SC, $x$=0.19, to the lower SC dopings suggests that there is a dramatic change in the normal ground state in LCCO at the end of the SC dome ($x_c$). Evidence for an anomalous critical behavior at $x_c$ has been reported [28] but not yet further investigated.

## Summary

We have discovered an unexpected behavior of the low temperature thermopower (S/T $\propto$ ln(1/T)) in the normal state of the electron-doped cuprate La$_{2-x}$Ce$_x$CuO$_4$ (LCCO) over an extended doping regime ($x$) above the Fermi surface reconstruction at $x$ =0.14. This suggests an anomalous quantum critical behavior in this system. Significantly, the magnitude of the slope of the logarithmic-in-T thermopower scales with the superconducting Tc, with both going to zero at the end of the superconducting dome. This suggests an intimate link between the quantum critical fluctuations and the Cooper pairing. We find a similar behaviour in another n-type cuprate, PCCO, strongly indicating that this is a universal behaviour in the electron-doped cuprates.

**Acknowledgement:** This work is supported by the NSF under Grant No.DMR-1708334 and the Maryland "Center for Nanophysics and Advanced Materials (CNAM).**References:**

1. Keimer B, et al. (2015) From quantum matter to high-temperature superconductivity in copper oxides. *Nature* 518:179.

2. Sachdev S, et al. (2011) Quantum criticality. *Phys. Today* 64(2): 29.

3. Yildirim Y, et al. (2015) Weak phase stiffness and nature of the quantum critical point in underdoped cuprates *Phys. Rev. B* 92:180501(R).

4. Sachdev S, (2003) Colloquium: Order and quantum phase transitions in the cuprate superconductors *Rev. Mod. Phys.* 75:913.

5. Badoux S, et al. (2016) Change of carrier density at the pseudogap critical point of a cuprate superconductor *Nature* (London) 531:210.

6. Tallon JL, et al. (2001) The doping dependence of $T^*$ – what is the real high-$T_c$ phase diagram?. *Physica C* 349:53.

7. Ramshaw BJ, et al. (2015) Quasiparticle mass enhancement approaching optimal doping in a high-$T_c$ superconductor. *Science* 348:317.

8. Mandal PR, et al. (2018) Nernst effect in the electron-doped cuprate superconductor $La_{2-x}Ce_xCuO_4$. *Phys. Rev. B* 97:014522.

9. Armitage NP, et al. (2010) Progress and perspectives on electron-doped cuprates. *Rev. Mod. Phys.* 82:2421.

10. Dagan Y, et al. (2004) Evidence for a Quantum Phase Transition in $Pr_{2-x}Ce_xCuO_{4-\delta}$ from Transport Measurements. *Phys. Rev. Lett.* 92:167001.7

**Figures**

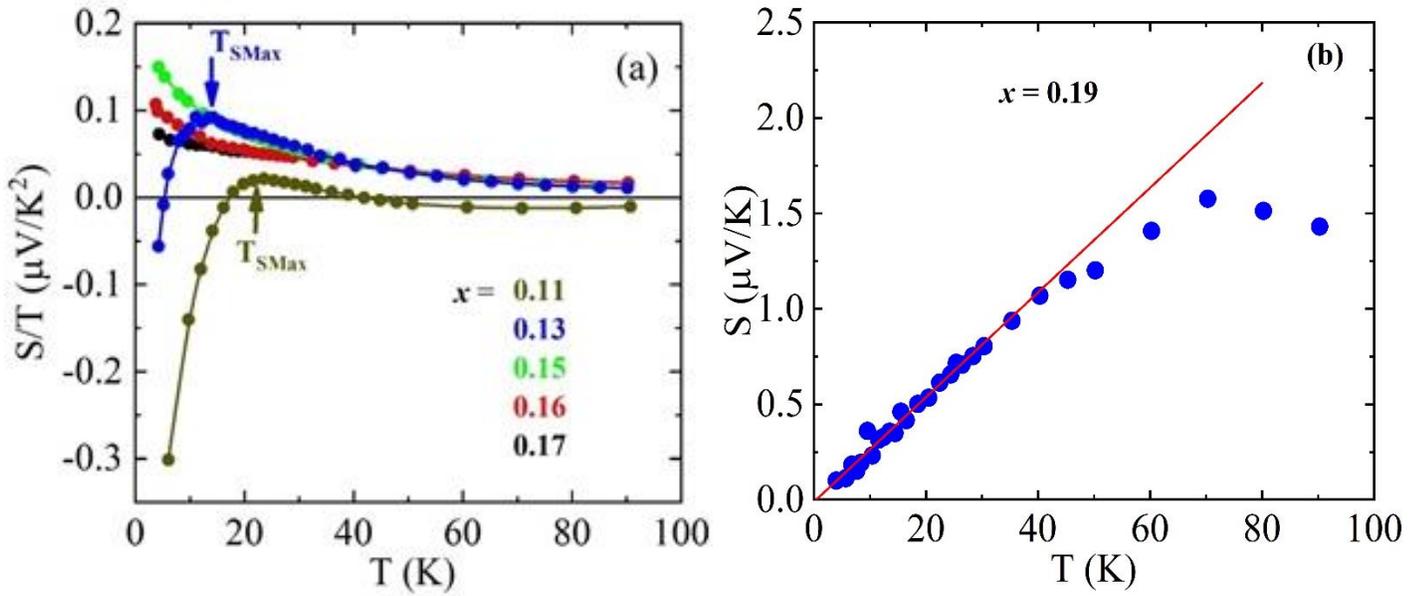

**Figure 1**: (a) Seebeck coefficient (S) of LCCO for different concentrations, plotted as S/T versus temperature $T$, measured at a magnetic field of 11 T for $x = 0.11$ to 0.17. $T_{Smax}$ denotes the temperature below which S/T decreases at low temperatures reaching negative values for $x = 0.11$ and 0.13. For $x = 0.11$ and 0.13 S/T data decrease below 26.5 and 13 K, respectively. For 0.15, 0.16, and 0.17 S/T data shows increasing behavior at low temperature. (b) S versus T for overdoped LCCO, $x = 0.19$ at zero field. The solid line is the fit to $S \propto T$ down to lowest measuring temperature, 4 K (see text).

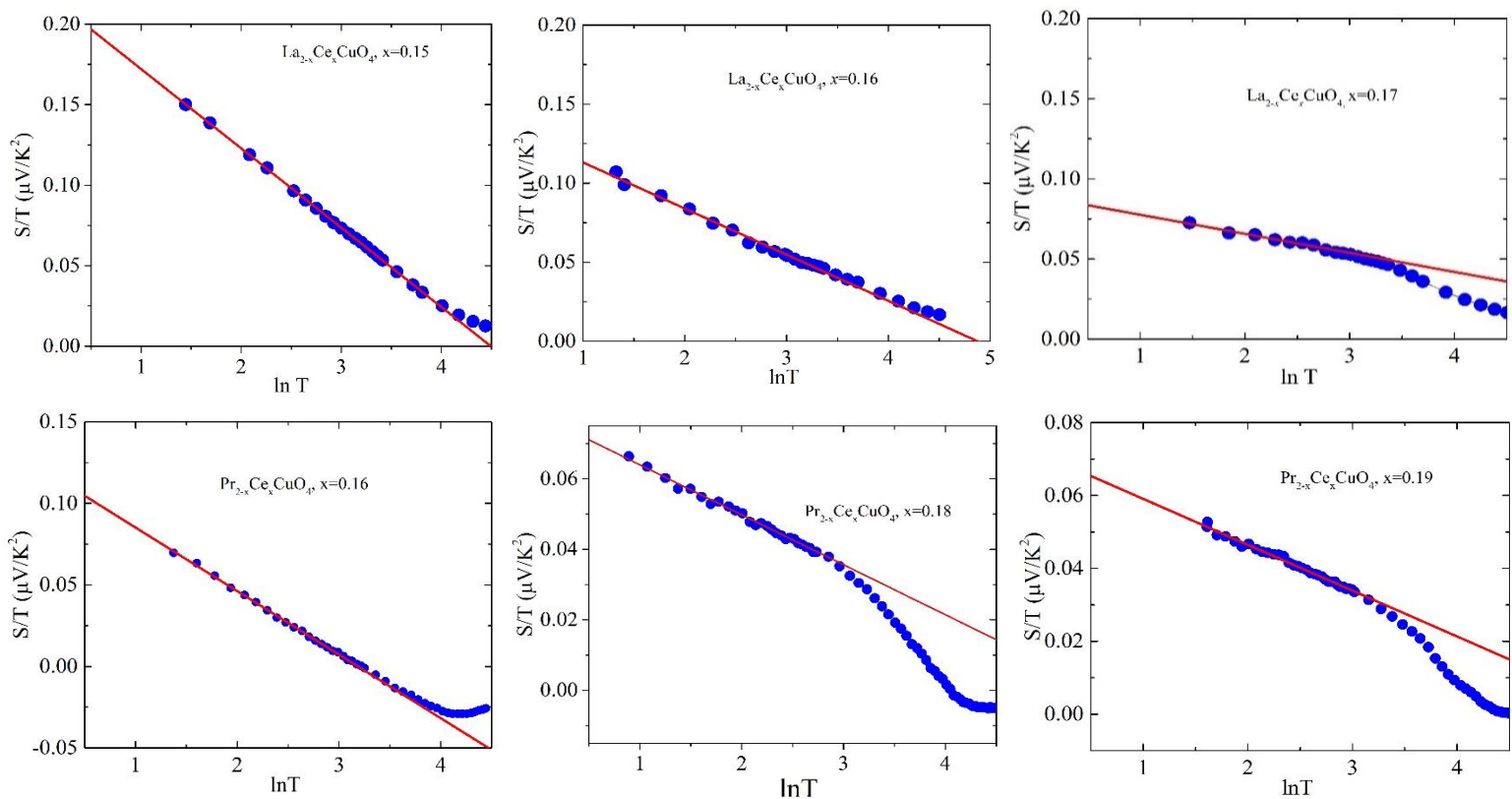

**Figure 2**: Normal state Seebeck coefficient for LCCO films $x \geq 0.14$ and PCCO films with $x \geq 0.16$, plotted as $S/T$ vs $\ln T$. The solid lines are a linear fit to the data down to lower temperature. For all the films $S/T$ exhibits $-\ln T$ dependence down to the lowest measured temperature of 4 K for LCCO and 3 K for PCCO.

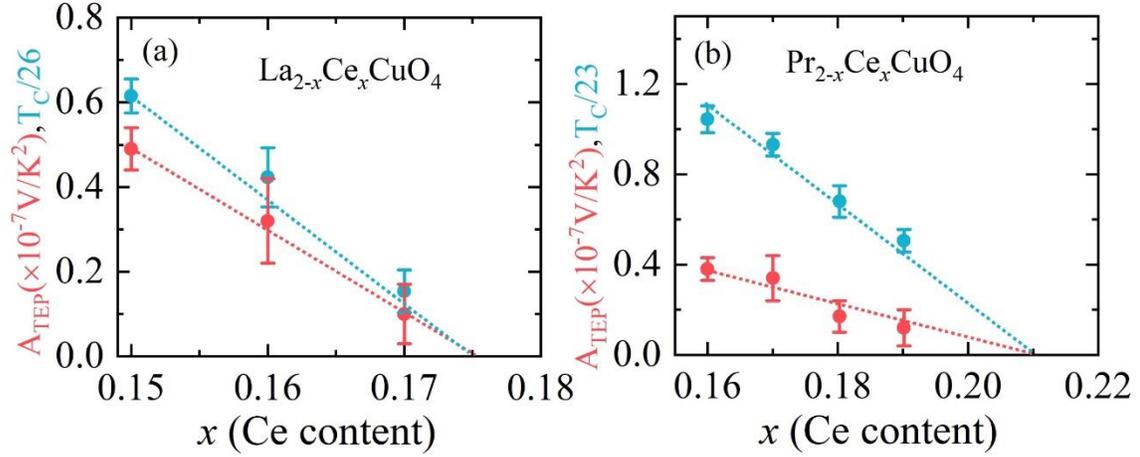

**Figure 3**: Doping dependence of $A_{TEP}$ (slope of the data in Fig. 2). Figures (a) and (b) show $A_{TEP}$ (red circles) and $T_C$ (divided by the superconducting transition temperature of the optimal doped 0.11 sample, 26 K for LCCO and 23 K for PCCO; blue circles) for different dopings of LCCO and PCCO. The error bars in $T_C$ are the standard deviation over many samples of each doping. The error bars in $A_{TEP}$ are a convolution of standard deviations in the values of the slopes for different temperature ranges of fitting.

# Supplementary material for "Anomalous quantum criticality in the electron-doped cuprates"

*Sample preparation*

The measurements have been performed on *c*-axis oriented $La_{2-x}Ce_xCuO_4$ thin films for the optimally doped ($x$=0.11), and overdoped ($x$=0.13, 0.15, 0.16, and 0.17) compositions. The thin films were deposited on (100) $SrTiO_3$ (10×5 mm$^2$) substrates by a Pulsed Laser deposition (PLD) technique utilizing a KrF excimer laser as the exciting light source [8] at a temperature of 700 °C and at an oxygen partial pressure of 230 mTorr. The thickness of the films used for this study is typically between 150 to 200 nm. The quality of the films was determined by the lowest residual resistivity of the samples and the superconducting transition width ($\Delta Tc$) calculated from the imaginary part of the AC susceptibility peak. The targets of the compounds for the PLD were prepared by the solid-state reaction method using 99.999% pure $La_2O_5$, $CeO_5$, and $CuO$ powders.

*Thermopower measurements technique*

The thermopower has been measured using two heater-two thermometer technique [25]. It is a steady state method where the temperature gradient is reversed to cancel out the Nernst effect and other possible background contributions. The sample is mounted on two copper blocks which are thermally insulated from a temperature controlled base. Two small chip resistor heaters are attached to the copper blocks and two tiny Lakeshore Cernox bare chip thermometers are on the two ends of the sample to monitor the temperature gradient (0.7-1 K) continuously. The temperature gradient is developed by applying power to each heater and the temperature gradient direction is switched by turning on or off the heaters. The electric voltage was measured using a Keithley 2001 multimeter with a sensitivity of several nanovolts in a Physical Property

Measurement System (PPMS) when the gradient was stable. The data were averaged for many times to reduce the random error. To reduce the contribution from the voltage leads we have used phosphor bronze wire which has a small thermopower even at high field in the temperature range of our measurements [29]. The measurements were performed under high vacuum and the magnetic field was applied perpendicular to the *ab* plane of the films. The sample temperature is taken as the average of hot and cold end temperatures. The Hall coefficient for the film $x$=0.13 was measured by applying a magnetic field (14 T) perpendicular to the film plane in the PPMS and averaging up and down field directions to eliminate any ab-plane magnetoresistive component.

*Hall coefficient and thermopower*

Figure S1 shows the Hall coefficient, $R_H$ (T) of LCCO film as a function of temperature with $x = 0.13$ at an applied magnetic field of 14 T measured from 100 to 1.8 K. The Hall coefficient is found to drop below T $\simeq$ 15 K and becomes negative below T $\simeq$ 8 K. The Hall coefficient for $x$=0.11 shows similar behavior and is also negative at lower temperature. The data are in good agreement with prior data for $x$=0.11 and 0.13 [18]. This drop is closely linked to the onset of spin density wave order. $T_{R_{Hmax}}$ is the temperature below which Hall coefficient starts to fall at lower temperature to reach negative values (arrow). This is another signature of the FSR. The prior Hall data of LCCO as a function of doping shows that the normal-state Hall coefficient suddenly drops and changes sign between 0.13 and 0.14 [18]. The phase diagram in Fig. S2 shows the doping evolution of $T_{Smax}$ (obtained from Fig. 1 (a)) with their uncertainty. Extrapolating this trend yields $T_{Smax}$ =0 at $x$=0.14. According to the Hall Effect studies at 400 mK, $La_{2-x}Ce_xCuO_4$ with $x$=0.14 has been suggested to be the critical doping where the normal state Fermi surface is known to undergo a reconstruction [18]. Figure S2 displays the temperature vs Ce concentration ($x$) phase diagram of $La_{2-x}Ce_xCuO_4$. The yellow regime presents the superconducting dome. Below $x$<0.14, the normal state in-plane resistivity shows a minima $T_{\rho min}$, which increases with decrease in $x$ [18] as shown in Fig. S2. The solid blue line is the estimated Fermi surface reconstruction line which separates the large FS from the reconstructed FS as a function of doping.

**Reference**

29. Y. Wang *et al.*, Nature, (London) **423**, 425 (2003).

**Figures**

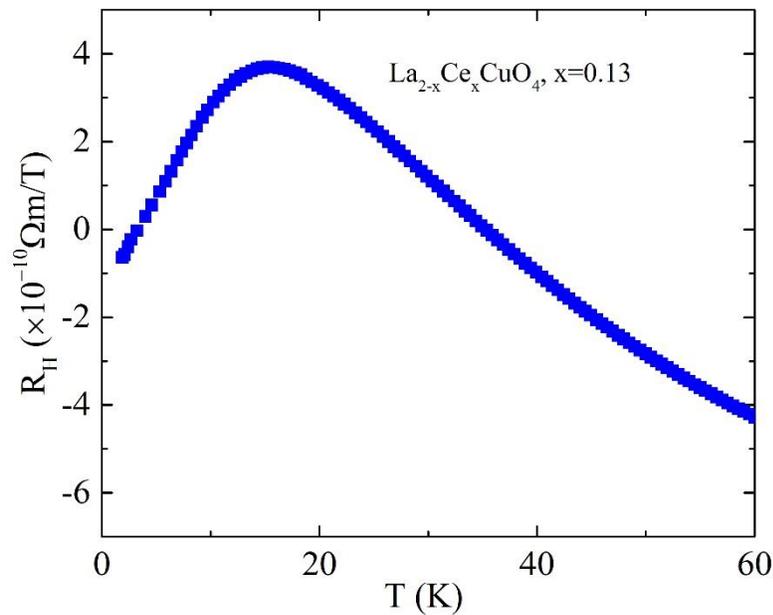

**Figure S1:** Hall coefficient versus temperature for $La_{2-x}Ce_xCuO_4$ films with $x$=0.13 measured at a magnetic field of 11 T.

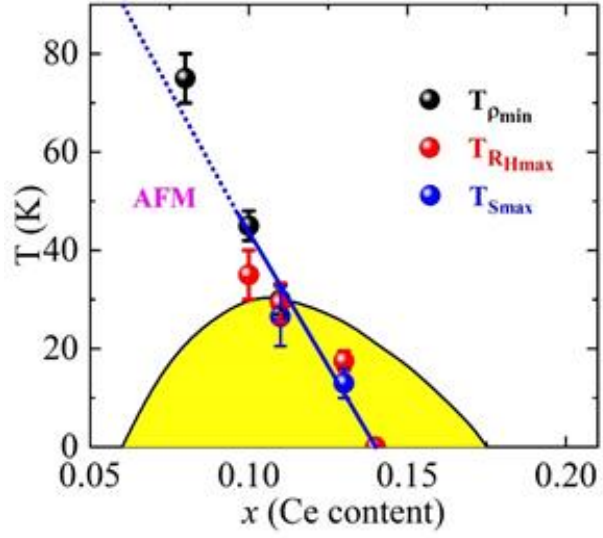

**Figure S2:** Temperature vs doping phase diagram of $La_{2-x}Ce_xCuO_4$. The black line denotes the superconducting transition temperature as a function of concentration and the yellow region denotes the superconducting dome. $T_{\rho min}$ and $T_{RHmax}$ are the normal-state in-plane resistivity minima and normal-state in-plane hall resistivity maxima, respectively [19]. Solid blue line (ending at $x = 0.14$) represents the FSR line separating the large Fermi surface from the reconstructed Fermi surface. The nearly similar value of $T_{RHmax}$, and $T_{Smax}$ for $x= 0.11$ and $0.13$ samples are the evidence of FSR.